\documentclass[preprint, 10pt, amsmath,amssymb, aps, prd]{revtex4-2}
\usepackage[utf8]{inputenc}
\usepackage{hyperref}
\usepackage{slashed}
\usepackage{color}
\usepackage{graphicx}
\graphicspath{{pics/}}
\allowdisplaybreaks[1]
\usepackage{multirow}

\begin{document}
\title{Dipole polarizabilities of light pseudoscalar mesons within the Domain Model of QCD vacuum}
\author{Sergei Nedelko}
\email{nedelko@theor.jinr.ru}
\author{Vladimir Voronin}
\email{voronin@theor.jinr.ru}
\affiliation{Joint Institute for Nuclear Research, 141980 Dubna, Moscow Region, Russia}
\begin{abstract}
Dipole polarizabilities of light pseudoscalar mesons are calculated in the framework of 
the mean field approach to QCD vacuum and bosonization based on the statistical ensemble 
of almost everywhere homogeneous Abelian (anti-)self-dual gluon fields, the domain model 
of QCD vacuum. In this approach, a nonlocal effective action of meson fields is derived 
which describes all possible strong, weak and 
electromagnetic interactions of meson fields including their excited states. 
The considered mean field implements confinement and chiral symmetry, which manifests 
itself both in the properties of quark and gluon fields, as well as, upon bosonization, 
in the mass spectrum, decay constants and form factors of nonlocal colorless hadrons, and 
leads to the qualitatively distinctive features of the effective meson action. Particularly 
relevant to the subject of the present paper are the nonlocality of meson-quark-antiquark vertices 
and  the absence of poles at real momenta in the propagators of scalar meson fields composed 
of light quark-antiquark pairs. 
In view of this, studying the role of manifest nonlocality of mesons     and  contribution 
of intermediate scalar meson fields in formation of the  polarizabilities  is of special interest.  
It turns out  that for charged pions and kaons, this contribution is substantial, but not the largest one. Nonlocal nature of mesons provides additional contribution, so that calculated polarizabilities are in reasonable agreement with COMPASS experimental data and Chiral Perturbation Theory.
\end{abstract}
\maketitle
\section{Introduction}
Polarizabilities of hadrons characterize their response to applied electromagnetic field which cannot be attributed to point-like particles, and are of fundamental interest for low-energy QCD. Experimental measurement of polarizabilities is challenging, so most data are available for lightest mesons $\pi^\pm,\pi^0$, but there is long-standing discrepancy in the values (see papers \cite{Ivanov:2015ntu,Moinester:2019sew,Moinester:2022tba} for review of theoretical and experimental status of the meson polarizability problem). Among the reported experimental results, only the most recent data on charged pion polarizability by COMPASS collaboration at CERN~\cite{COMPASS:2014eqi} is consistent with Chiral Perturbation Theory (ChPT). The leading-order result of ChPT~\cite{Bijnens:1987dc} is equivalent to the value found in Ref.~\cite{Terentev:1972ix} based on the hypothesis of partially conserved axial-vector current.

The polarizabilities were investigated theoretically within Chiral Perturbation Theory up to two loops~\cite{Gasser:2006qa,Burgi:1996qi,Burgi:1996mm,Bellucci:1994eb,Gasser:2005ud}, with the methods of Lattice QCD~\cite{Fiebig:1988en,Lee:2005dq,Lujan:2014kia,Freeman:2014kka,Luschevskaya:2015cko,Lujan:2016ffj,Bali:2017ian,Niyazi:2021jrz,Bignell:2020dze,Ding:2020hxw,Wilcox:2021rtt} , within various phenomenological models~\cite{Lvov:1980st,Volkov:1980cj,Volkov:1985nk,Bernard:1988gp,Ivanov:1991kw,Dorokhov:1997rv,Donoghue:1993kw,Hiller:2009ik} and with the help of dispersion relations~\cite{Filkov:1982cx,Filkov:1998rwz,Filkov:2005suj}. Several studies~\cite{Ivanov:1991kw,Filkov:1998rwz,Dorokhov:1997rv,Hiller:2009ik} found that dominating part of pion polarizabilities is due to $\sigma$ meson. In the present study, polarizabilities are extracted from the nonlocal effective meson action deduced within the Domain Model of QCD vacuum and hadronization (see Refs.~\cite{Kalloniatis:2003sa,Nedelko:2016gdk,Nedelko:2016vpj,Nedelko:2014sla,Nedelko:2020bba}).

In this action meson fields appear as collective colorless excitations of confined dynamical quark-antiquark, heavy and light ones,  and gluon fields. A highly nonlocal effective meson action contains information about the  strong, weak and electromagnetic interactions of mesons as well as their two-point correlation functions.    In particular, the model systematically describes various phenomena related to confinement and chiral symmetry realization, the heavy quark limit.   Meson masses, including Regge spectrum of excited states of mesons, their decay constants and form factors  are in good agreement with experimental values~\cite{Efimov:1995uz,Burdanov:1996uw,Kalloniatis:2003sa,Nedelko:2016gdk,Nedelko:2016vpj,Nedelko:2014sla,Nedelko:2020bba}.

The specific feature of this approach is that mesons appear as extended composite fields due to nonlocal meson-quark vertices. Meson masses are identified as the poles of the nonlocal meson propagators.  As it turns out, there is no poles at real momenta in the nonlocal propagators of the scalar meson-like composite fields, and therefore light  scalar mesons as  quark-antiquark states are absent in the physical spectrum of the stable collective excitations.  This property relates to the peculiarities of realization of chiral symmetry in the presence of Abelian (anti)-self-dual gluon mean field and agrees with expectation that the lightest scalar state $\sigma$ is not intuitively made of a quark and an antiquark~\cite{Pelaez:2015qba}. At the same time,  scalar meson-like fields contribute to the amplitudes of various processes.  In has to be noted that physical scalar states occur in the hyperfine splitting of the orbital excitations of the vector mesons with the masses above 1 GeV.

We describe the formalism and  its features in Section~\ref{section_model}.
In Section~\ref{section_polarizabilities} we calculate polarizabilities and find that charged pion polarizability is consistent with ChPT~\cite{Gasser:2006qa} and COMPASS data~\cite{COMPASS:2014eqi}. The contribution of intermediate scalar meson fields to polarizability of pion and kaon is substantial,  especially for the neutral ones, while for charged pion and kaon polarizability  intermediate scalar meson turns out to be less  important.

\section{Effective meson action\label{section_model}}

The mean-field approach based on the random Abelian (anti-)self-dual vacuum gluon fields allows to deduce a generating functional via bosonization of one-gluon exchange interaction of quark currents which has the form~\cite{Efimov:1995uz,Kalloniatis:2003sa,Nedelko:2016gdk,Nedelko:2016vpj,Nedelko:2014sla,Nedelko:2020bba}:
\begin{align}
\label{meson_pf}
Z&={\cal N}
\int D\phi_{\cal Q}
\exp\left\{-\frac{\Lambda^2}{2}\frac{h^2_{\cal Q}}{g^2 C^2_\mathcal{Q}}\int d^4x 
\phi^2_{\cal Q}(x)
-\sum\limits_{k=2}^\infty\frac{1}{k}W_k[\phi]\right\},
\\
\label{effective_meson_action}
W_k[\phi]&=
\sum\limits_{{\cal Q}_1\dots{\cal Q}_k}h_{{\cal Q}_1}\dots h_{{\cal Q}_k}
\int d^4x_1\dots\int d^4x_k
\Phi_{{\cal Q}_1}(x_1)\dots \Phi_{{\cal Q}_k}(x_k)
\Gamma^{(k)}_{{\cal Q}_1\dots{\cal Q}_k}(x_1,\dots,x_k),
\\
\label{physical_meson_fields}
\Phi_{{\cal Q}}(x)&=\int \frac{d^4p}{(2\pi)^4}e^{ipx}{\mathcal O}_{{\mathcal Q}{\mathcal Q}'}(p)\tilde\phi_{{\mathcal Q}'}(p),\quad C_\mathcal{Q}=C_J,\quad C^2_{S/P}=2C^2_{V/A}=\frac{1}{9}.
\end{align}
The condensed index $\mathcal{Q}\equiv\{aJLn\}$ includes all quantum numbers of a meson, $\Lambda$ is a scale  related to the strength of the vacuum gluon field, and  finally  to the value of gluon condensate $\langle g^2 F^2 \rangle$.  The physical color neutral meson fields $\phi_{{\mathcal Q}}$ are obtained by means of  orthogonal transformation ${\mathcal O}_{{\mathcal Q}{\mathcal Q}'}$ of fields $\Phi_{{\mathcal Q}}$.   The quadratic part of the action for $\phi_{{\mathcal Q}}$ is diagonal with respect to all quantum numbers. The masses of mesons correspond to the poles of nonlocal propagators
\begin{equation}
\label{meson_propagator}
D_\mathcal{Q}(p^2)=h_\mathcal{Q}^{-2} \left(\frac{\Lambda^2}{g^2 C^2_\mathcal{Q}}+\tilde\Gamma^{(2)}_{\mathcal{Q}}(p^2)\right)^{-1}
\end{equation}
and can be found as zeroes of inverse propagator from equation
\begin{align}
\label{mass-eq}
0&=
\frac{\Lambda^2}{g^2 C^2_\mathcal{Q}}+\tilde\Gamma^{(2)}_{\cal Q}(-M^2_{\cal Q}),
\end{align}
where $\tilde\Gamma^{(2)}_{\cal Q}$ is two-point correlation function diagonalized with respect to all quantum numbers. 
Constants $h_\mathcal{Q}$ are defined by equation
\begin{align}
\nonumber 
1&=h^2_{\cal Q}
\left.\frac{d}{dp^2}\tilde\Gamma^{(2)}_{\cal Q}(p^2)\right|_{p^2=-M^2_{\cal Q}}
\end{align}
which ensure that residue at the pole of the propagator is equal to unity. The results of calculation of the masses of various mesons as well as analytical expressions of $\tilde\Gamma^{(2)}_{\cal Q}(p^2)$ can be found in Ref.~\cite{Nedelko:2016gdk}. In the one-loop approximation, the meson propagators given by Eq.~\eqref{meson_propagator} are real, and therefore mesons are stable with respect to decay into quarks by virtue of the optical theorem. It is also quite plausible that correlation functions of greater number of external mesons are suppressed as number of colors $N_c$ approaches infinity because the functional~\eqref{meson_pf} is deduced from QCD. Thorough investigation of this limit is an interesting topic to be studied in the future.

Correlation functions include ``connected'' and ``disconnected'' contributions of quark loops in the background field. For example, the two-point nonlocal vertex function $\tilde\Gamma^{(2)}_{\cal QQ'}(p)$ is given by
\begin{equation}
\label{Gammak}
\Gamma^{(2)}_{{\cal Q}_1{\cal Q}_2}=
\overline{G^{(2)}_{{\cal Q}_1{\cal Q}_2}(x_1,x_2)}-
\Xi_2(x_1-x_2)\overline{G^{(1)}_{{\cal Q}_1}G^{(1)}_{{\cal Q}_2}},
\end{equation}
where $\Xi$ is correlation function of the background field which belongs to the statistical ensemble of the almost everywhere homogeneous Abelian (anti-)self-dual fields.
Quark loops $G^{(k)}_{{\cal Q}_1\dots{\cal Q}_k}$ are averaged over the ensemble of background field configurations with measure $d\sigma_B$:
\begin{eqnarray}
\label{barG}
&&\overline{G^{(k)}_{{\cal Q}_1\dots{\cal Q}_k}(x_1,\dots,x_k)}
=\int d \sigma_B
{\rm Tr}V_{{\cal Q}_1}\left(x_1\right)S\left(x_1,x_2\right)\dots
V_{{\cal Q}_k}\left(x_k\right)S\left(x_k,x_1\right),
\\
&&\overline{G^{(l)}_{{\cal Q}_1\dots{\cal Q}_l}(x_1,\dots,x_l)
G^{(k)}_{{\cal Q}_{l+1}\dots{\cal Q}_k}(x_{l+1},\dots,x_k)}
=
\nonumber\\
\nonumber
&&\int d \sigma_B
{\rm Tr}\left\{
V_{{\cal Q}_1}\left(x_1\right)S\left(x_1,x_2\right)\dots
V_{{\cal Q}_k}\left(x_l\right)S\left(x_l,x_1\right)
\right\}\times
\\
&&
{\rm Tr}\left\{
V_{{\cal Q}_{l+1}}\left(x_{l+1}\right)S\left(x_{l+1},x_{l+2}\right)\dots
V_{{\cal Q}_k}\left(x_k\right)S\left(x_k,x_{l+1}\right)
\right\}.
\nonumber 
\end{eqnarray}
Here $S(x,y)$ is the quark propagator and $V_{{\cal Q}}$ are  nonlocal  meson-quark-antiquark vertices. 

The quark propagator and meson-quark vertices in the presence of the almost everywhere homogeneous fields are approximated by those in the homogeneous (anti-)self-dual Abelian background field. The averaging over mean field ensemble is achieved by averaging the quark loops over configurations of the homogeneous  background fields, supplemented by taking into account  $n$-point correlators of the mean fields $\Xi_n$. The averaging is performed over self-dual and anti-self-dual Abelian (anti-)self-dual configurations and their  directions in Euclidean and color spaces. Averaging over spatial directions in $R^4$ is performed with the help of generating formula
\begin{equation}
\label{averaging_over_vacuum_field}
\int d\sigma_B\exp(if_{\mu\nu}J_{\mu\nu})=\langle\exp(if_{\mu\nu}J_{\mu\nu})\rangle=\frac{\sin\sqrt{2\left(J_{\mu\nu}J_{\mu\nu}\pm J_{\mu\nu}\widetilde{J}_{\mu\nu}\right)}}{\sqrt{2\left(J_{\mu\nu}J_{\mu\nu}\pm J_{\mu\nu}\widetilde{J}_{\mu\nu}\right)}},
\end{equation}
where $J_{\mu\nu}$ is an arbitrary antisymmetric tensor.
Tensor $f_{\mu\nu}$ is an appropriately normalized Abelian (anti-)self-dual background field with strength $\Lambda$:
\begin{gather}
\nonumber
\hat B_\mu=-\frac{1}{2}\hat n B_{\mu\nu}x_\nu, \ \hat n = t^3\cos\xi+t^8\sin\xi,
\\
\label{b-field}
\tilde{B}_{\mu\nu}=\frac12\epsilon_{\mu\nu\alpha\beta}B_{\alpha\beta}=\pm B_{\mu\nu}, \  \hat{B}_{\rho\mu}\hat{B}_{\rho\nu}=4\upsilon^2\Lambda^4\delta_{\mu\nu},\\
\nonumber
f_{\alpha\beta}=\frac{\hat{n}}{2\upsilon\Lambda^2}B_{\alpha\beta}, \  \upsilon=\mathrm{diag}\left(\frac16,\frac16,\frac13\right), \ f_{\mu\alpha}f_{\nu\alpha}=\delta_{\mu\nu},
\end{gather}
where the upper sign in ``$\pm$'' should be taken for self-dual field, and the lower for anti-self-dual field.
Nonlocal vertices $V^{aJln}_{\mu_1\dots\mu_l}$ are given by formulas
\begin{gather}
V^{aJln}_{\mu_1\dots\mu_l}= {\cal C}_{ln}\mathcal{M}^a\Gamma^J F_{nl}\left(\frac{\stackrel{\leftrightarrow}{\cal D}^2\!\!\!
(x)}{\Lambda^2}\right)T^{(l)}_{\mu_1\dots\mu_l}\left(\frac{1}{i}\frac{\stackrel{\leftrightarrow}{\cal D}\!(x)}{\Lambda}\right),
\label{qmvert}\\
{\cal C}^2_{ln}=\frac{l+1}{2^ln!(n+l)!},\quad F_{nl}(s)=s^n\int_0^1 dt t^{n+l} \exp(st),
\nonumber\\
{\stackrel{\leftrightarrow}{\mathcal{D}}}\vphantom{D}^{ff'}_{\mu}=\xi_f\stackrel{\leftarrow}{\mathcal{D}}_{\mu}-\ \xi_{f'}\stackrel{\rightarrow}{\mathcal{D}}_{\mu}, 
\ \
\stackrel{\leftarrow}{\mathcal{D}}_{\mu}\hspace*{-0.3em}(x)=\stackrel{\leftarrow}{\partial}_\mu+\ i\hat B_\mu(x),  \ \ 
\stackrel{\rightarrow}{\mathcal{D}}_{\mu}\hspace*{-0.3em}(x)=\stackrel{\rightarrow}{\partial}_\mu-\ i\hat B_\mu(x), 
\nonumber\\
\xi_f=\frac{m_{f'}}{m_f+m_{f'}},\ \xi_{f'}=\frac{m_{f}}{m_f+m_{f'}}.
\nonumber
\end{gather}
Here $\mathcal{M}^a$ and $\Gamma^J$ are flavor and Dirac matrices corresponding to a given meson field, $\xi_f,\xi_{f'}$ provide that $x$ is the center of mass of a meson, $n,l$ are radial and orbital quantum numbers, respectively.
Radial part  $F_{nl}$ is defined by the propagator of the gluon fluctuations  charged with respect to the Abelian background, $T^{(l)}$ are irreducible tensors of four-dimensional rotation group.
Propagator of the quark with mass $m_f$ in the presence of the homogeneous Abelian (anti-)self-dual field  has the form 
\begin{align}
\label{quark_propagator}
S_f(x,y)&=\exp\left(-\frac{i}{2}\hat n x_\mu  B_{\mu\nu}y_\nu\right)H_f(x-y),
\\
\tilde H_f(p)&=\frac{1}{2\upsilon \Lambda^2} \int_0^1 ds e^{(-p^2/2\upsilon \Lambda^2)s}\left(\frac{1-s}{1+s}\right)^{m_f^2/4\upsilon \Lambda^2}
\nonumber\\
&\quad\times \left[\vphantom{\frac{s}{1-s^2}}p_\alpha\gamma_\alpha\pm is\gamma_5\gamma_\alpha f_{\alpha\beta} p_\beta
+m_f\left(P_\pm+P_\mp\frac{1+s^2}{1-s^2}-\frac{i}{2}\gamma_\alpha f_{\alpha\beta}\gamma_\beta\frac{s}{1-s^2}\right)\right],
\nonumber
\end{align}
where anti-Hermitean representation of Dirac matrices is used, and ``$\pm$'' signs are arranged in accordance with formula~\eqref{b-field}. The translation-invariant part $H_f$ of the propagator is an analytical function in the finite complex momentum plane and matches the behavior of free Dirac propagator at large Euclidean momentum. The analyticity of quark propagator is interpreted as confinement of dynamical quarks.

Overall, the mass spectrum of the  ground state and excited mesons composed of light and heavy quarks is described rather accurately, in complete agreement with
expectations based on confinement (Regge mass spectrum of radially and orbitally excited states) and chiral symmetry breaking (light pseudoscalar and heavy vector nonets, etc.) as well as asymptotic heavy-quark relations.  

\begin{figure}
\includegraphics[scale=1]{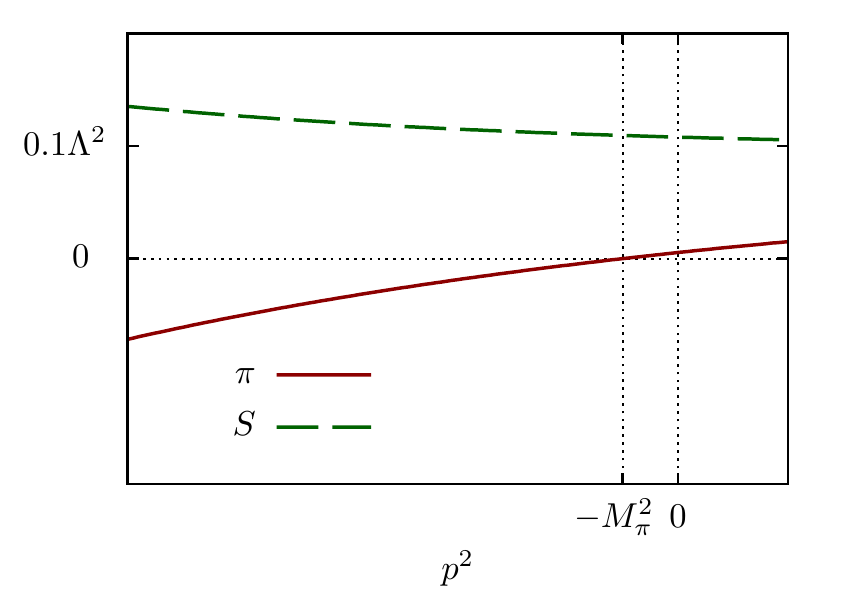}
\caption{Left-hand side of Eq.~\ref{mass-eq} (which is proportional to inverse propagator) for pion and ground-state scalar quark-antiquark field with respect to Euclidean momenta $p^2$. The zero of inverse pion propagator (correspondingly, the pole of propagator $D_\pi(p^2)$) is located at $-M_\pi^2$, while inverse propagator $D_S^{-1}(p^2)$ has no zeroes at real momenta.\label{figure_pi_propagator}}
\end{figure}

A peculiar property of the quark propagator in the mean gluon field under consideration is that, unlike the case of the pseudoscalar and vector ground-state mesons, there are no real solutions to equation~\eqref{mass-eq} for their parity partners, the  ground-state scalar and axial mesons. Axial and scalar mesons appear with a mass above $1$ GeV in the hyperfine splitting of the orbital excitations. The inverse propagators of pseudoscalar and scalar meson fields are shown in Fig.~\ref{figure_pi_propagator} which manifestly illustrates the absence of a pole for the scalar field propagator.   This feature
 is particularly relevant to the present study. Contributions of intermediate scalar meson-like fields to various processes are available and can be computed  but  a controversial  issue of  existence  of the light scalar mesons does not occur.

Though the meson-quark coupling constant  $h_\mathcal{Q}$ is obviously undefined for meson-like composite fields if corresponding Eq.~\eqref{mass-eq} has no solutions, it is convenient to keep universal notation. Such fields can only be virtual, and $h_\mathcal{Q}$ cancels out in final expressions ($h^2_\mathcal{Q}$ for two vertices cancel $h_\mathcal{Q}^{-2}$ in propagator $D_\mathcal{Q}$).

Electromagnetic interactions are included in gauge-invariant way using the prescription of Ref.~\cite{Terning:1991yt} which yields expansions (see~\cite{Nedelko:2016gdk,Nedelko:2021dsh})
\begin{align}
\nonumber 
S_f(x,y|A)&=S_f(x,y)+\sum_{n=1}^\infty \left(Q_fe\right)^n \int dz_1\cdots \int dz_n S_f(x_1,z_1)\gamma_{\mu_1}A_{\mu_1}(z_1)\cdots S_f(z_{i-1},z_i)\gamma_{\mu_i}A_{\mu_i}(z_i)\cdots S_f(z_n,y),\\
\nonumber 
V_\mathcal{Q}(x|A)&=V_\mathcal{Q}(x)+\sum_{n=1}^\infty e^n \int dz_1\cdots \int dz_n V_{\mathcal{Q}\mu_1\dots\mu_n}(x,z_1,\dots,z_n)A_{\mu_1}(z_1)\cdots A_{\mu_n}(z_n),
\end{align}
where $Q$ is a diagonal matrix of quark charges in units of electron charge $e$, and meson-photon vertices appear due to nonlocality of meson-quark interactions. 
One-photon and two-photon meson vertices are given by
\begin{gather}
\label{1-photon-vertex}
V_{\mathcal{Q}\mu}(x;q)= \int_0^1 d\tau \frac{1}{\tau} \frac{\partial}{\partial q_\mu} \left\{-Q_f V_{\mathcal{Q}}\left( \stackrel{\leftrightarrow}{\mathcal{D}}(x) -iq\tau\xi\right)+ Q_{f'} V_{\mathcal{Q}}\left(\stackrel{\leftrightarrow}{\mathcal{D}}(x) + iq\tau\xi'\right)\right\},\\
\label{2-photon-vertex}
\begin{split}
V_{\mathcal{Q}\mu\nu}(x;q_1,q_2)=\frac{1}{2}\int_0^1 d\tau_1 \int_0^1 d\tau_2 \frac{1}{\tau_1 \tau_2}\frac{\partial}{\partial q_{1\mu}} \frac{\partial}{\partial q_{2\nu}}  
& \left[Q_f Q_{f'}V_\mathcal{Q}\left( \stackrel{\leftrightarrow}{\mathcal{D}}(x) -iq_1\tau_1\xi -iq_2\tau_2\xi\right)\right.
\\
&-Q_f Q_{f'} V_\mathcal{Q}\left( \stackrel{\leftrightarrow}{\mathcal{D}}(x) -iq_1\tau_1\xi +iq_2\tau_2\xi\right)\\
&-Q_{f'} Q_f V_\mathcal{Q}\left( \stackrel{\leftrightarrow}{\mathcal{D}}(x) +iq_1\tau_1\xi -iq_2\tau_2\xi\right)\\
&\left.+Q_{f'} Q_{f'} V_\mathcal{Q}\left( \stackrel{\leftrightarrow}{\mathcal{D}}(x) +iq_1\tau_1\xi +iq_2\tau_2\xi\right)\right],
\end{split}
\end{gather}
where $Q_f$ is electric  charge of a quark with flavor $f$.

The generating functional and the effective meson action take the form
\begin{align}
\label{meson_pf_em}
Z&={\cal N}
\int D\phi_{\cal Q}\int DA_\mu
\exp\left\{-\frac{1}{4}\int d^4x\ F_{\mu\nu}F_{\mu\nu}-\frac{\Lambda^2}{2}\frac{h^2_{\cal Q}}{g^2 C^2_\mathcal{Q}}\int d^4x 
\phi^2_{\cal Q}(x)
-\Gamma(A)
-\sum\limits_{k=2}^\infty\frac{1}{k}W_k[\phi|A]\right\},
\\
\nonumber 
W_k[\phi|A]&=
\sum\limits_{{\cal Q}_1\dots{\cal Q}_k}h_{{\cal Q}_1}\dots h_{{\cal Q}_k}
\int d^4x_1\dots\int d^4x_k
\Phi_{{\cal Q}_1}(x_1)\dots \Phi_{{\cal Q}_k}(x_k)
\Gamma^{(k)}_{{\cal Q}_1\dots{\cal Q}_k}(x_1,\dots,x_k|A),
\end{align}
where
\begin{equation*}
\Gamma(A)=\int d\sigma_B \mathrm{Tr}\log\left[1+ Qe\gamma_\mu A_\mu(x)S(x,y)\right],
\end{equation*}
and
$\Gamma^{(k)}_{{\cal Q}_1\dots{\cal Q}_k}(x_1,\dots,x_k|A)$ are obtained from $\Gamma^{(k)}_{{\cal Q}_1\dots{\cal Q}_k}(x_1,\dots,x_k)$ by substitutions
\begin{equation*}
S_f(x,y)\to S_f(x,y|A),\quad
V_\mathcal{Q}(x)\to V_\mathcal{Q}(x|A).
\end{equation*}

The free  parameters of the model are scale $\Lambda$ (scalar gluon condensate), infrared limits of dynamical quark masses and strong coupling $\alpha_s$ which has been determined by fitting to the masses of mesons $\pi,\rho,K,K^*,J/\psi,\Upsilon,\eta'$ (for details see~\cite{Nedelko:2016gdk}). 

As it has been mentioned,  diagonalization of the quadratic part of the effective action~\eqref{effective_meson_action} with respect to the radial quantum number is a part of the calculation procedure. In practice some finite number of excited states can be taken into account.  
As it has been analyzed in~\cite{Nedelko:2016gdk}, though typically about five lowest radial states have to be taken into account for robust stability of the computation, a consistent overall description of the mass spectrum of mesons  is achieved irrespective to a number of accounted  radial excitations. Just the values of the free parameters  have to be adjusted when the number of accounted radial excitation changes. 
 
\section{Evaluation of polarizabilities\label{section_polarizabilities}}

The polarizabilities are defined by the Compton scattering amplitude
\begin{equation*}
P(p)+\gamma(q,\varepsilon)\to P(p')+\gamma(q',\varepsilon')
\end{equation*}
of a pseudoscalar meson $P$
\begin{equation*}
{}_\text{out}\langle P(p')\gamma(q',\varepsilon')|P(p)\gamma(q,\varepsilon) \rangle_\text{in}=i(2\pi)^4\delta^{(4)}(p'+q'-p-q)\varepsilon^\mu(q)\varepsilon^{*\nu}(q')M_{\mu\nu}.
\end{equation*}
We concentrate on the electric $\alpha_\text{E}$ and magnetic $\beta_\text{M}$ dipole polarizabilities which appear in expansion of the amplitude in small photon momenta as
\begin{equation*}
\varepsilon^\mu(q)\varepsilon^{*\nu}(q')M_{\mu\nu}=-2e^2\vec{\varepsilon}\cdot \vec{\varepsilon} {\,}'^{*}+8\pi M(\alpha_\text{E}\omega\omega'\,\vec{\varepsilon}\cdot \vec{\varepsilon} {\,}'^{*}+\beta_\text{M}\left(\vec{\varepsilon}\times \vec{q}\,\right)\cdot\left(\vec{\varepsilon}{\,}'^{*}\times \vec{q}{\,}'\right))+\dots,
\end{equation*}
where $M$ is the mass of a pseudoscalar meson.
The tensor $M_{\mu\nu}$ can be separated into two parts
\begin{equation}
\label{born_and_nonborn}
M_{\mu\nu}=M_{\mu\nu}^\text{Born}+M_{\mu\nu}^\text{NB},
\end{equation}
where the part $M_{\mu\nu}^\text{NB}$ describes the response of a meson as a composite system to the applied electromagnetic field. The term $M_{\mu\nu}^\text{Born}$ given by
\begin{equation}
\label{born_scattering}
M_{\mu\nu}^\text{Born}=e^2\left[2g_{\mu\nu}-\frac{(2p_\mu+q_\mu)(2p'_\nu+q'_\nu)}{(p+q)^2-M^2}-\frac{(2p_\nu-q'_\nu)(2p'_\mu-q_\mu)}{(p-q')^2-M^2}\right]
\end{equation}
describes real Compton scattering of a structureless pseudoscalar particle.

In the case of real Compton scattering ($q^2={q'}^2=0$, $\varepsilon^\mu(q) q_\mu=\varepsilon^{*\mu}(q')q'_\mu=0$), the tensor $\mathcal{M}_{\mu\nu}$ contains only two independent tensor structures~\cite{Tarrach:1975tu,Bardeen:1968ebo,Lvov:2001zdg}
\begin{align}
\label{compton_tensor_parametrization}
\varepsilon^\alpha(q)\varepsilon^{*\beta}(q')M_{\alpha\beta}&=\varepsilon^\alpha(q)\varepsilon^{*\beta}(q')\left(A T_{1\alpha\beta}+B T_{2\alpha\beta}\right),\\
\nonumber 
T_{1\alpha\beta}&=-\frac{t}{2}g_{\alpha\beta}-q_{\beta}q'_{\alpha},\\
\nonumber 
T_{2\alpha\beta}&=-\frac{t}{2}P_\alpha P_\beta+\nu^2 g_{\alpha\beta}-\nu(P_\alpha q_\beta+P_\beta q'_\alpha),  
\end{align}
where
\begin{equation*}
P=\frac{p+p'}{2},\quad s=(p+q)^2,\quad t=(q-q')^2,\quad u=(p-q')^2,\quad \nu=\frac{1}{4}(s-u).
\end{equation*}
In accordance with Eqs.~\eqref{born_and_nonborn} and~\eqref{born_scattering}, amplitudes $A$ and $B$ can be split in two parts
\begin{align*}
A(\nu^2,t)&=A^\text{Born}(\nu^2,t)+A^\text{NB}(\nu^2,t),\\
B(\nu^2,t)&=B^\text{Born}(\nu^2,t)+B^\text{NB}(\nu^2,t),
\end{align*}
where $A^\text{Born},B^\text{Born}$ are given by
\begin{equation*}
A^\text{Born}(\nu^2,t)=-\frac{e^2t}{(s-M_\pi^2)(u-M_\pi^2)},\quad B^\text{Born}(\nu^2,t)=-\frac{8e^2}{(s-M_\pi^2)(u-M_\pi^2)}.
\end{equation*}
The electric $\alpha_\text{E}$ and magnetic $\beta_\text{M}$ polarizabilities are related to $A^\text{NB},B^\text{NB}$ by means of equations (see \cite{Guiasu:1978dz,Guiasu:1979sz})
\begin{align}
\label{alpha_plus_beta_definition}
\alpha_\text{E}+\beta_\text{M}&=-\frac{M}{8\pi}B^\text{NB}(0,0),\\
\label{alpha_minus_beta_definition}
\alpha_\text{E}-\beta_\text{M}&=-\frac{1}{8\pi M}\left(2A^\text{NB}(0,0)+M^2B^\text{NB}(0,0)\right).
\end{align}

\begin{figure}
\begin{tabular}{c@{\hspace*{1em}}c@{\hspace*{1em}}c@{\hspace*{1em}}c}
\includegraphics[scale=1]{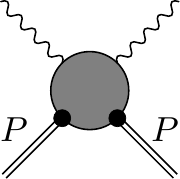}&
\includegraphics[scale=1]{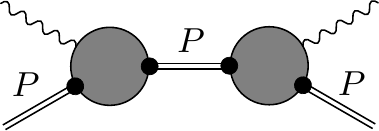}&
\includegraphics[scale=1]{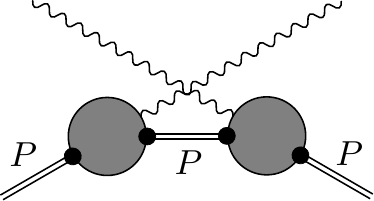}&
\includegraphics[scale=1]{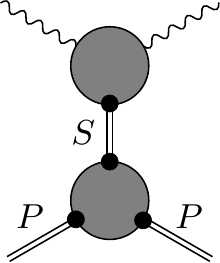}\\
(a)&(b)&(c)&(d)
\end{tabular}
\caption{
Diagrams contributing to Compton scattering tensor $M_{\mu\nu}$. The grey circles denote all possible one-particle irreducible contributions.
\label{compton_scattering_diagrams}}
\end{figure}

The diagrams that contribute to Compton tensor are shown in Fig.\ref{compton_scattering_diagrams}.
Consider contribution of diagrams (a),(b),(c) in Fig.\ref{compton_scattering_diagrams} which form gauge-invariant combination.
\begin{figure}
\begin{tabular}{c@{\hspace*{1em}}c@{\hspace*{1em}}c@{\hspace*{1em}}c@{\hspace*{1em}}c}
\includegraphics[scale=.75]{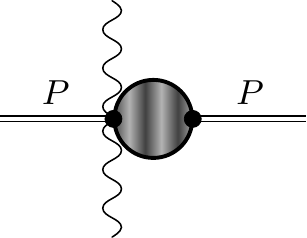}&
\includegraphics[scale=.75]{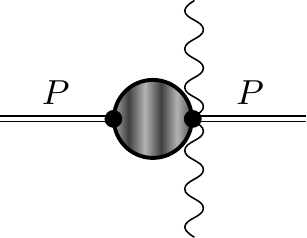}&
\includegraphics[scale=.75]{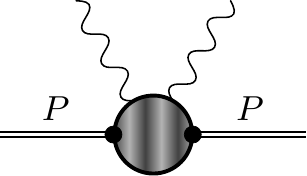}&
\includegraphics[scale=.75]{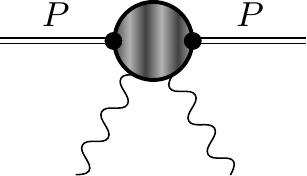}&
\includegraphics[scale=.75]{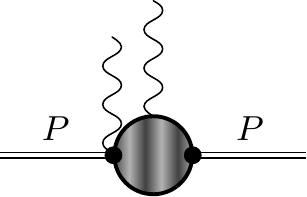}\\
(a)&(b)&(c)&(d)&(e)
\\[1ex]
\includegraphics[scale=.75]{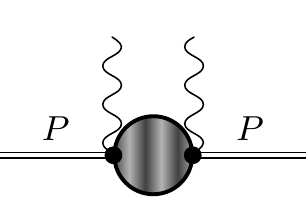}&
\includegraphics[scale=.75]{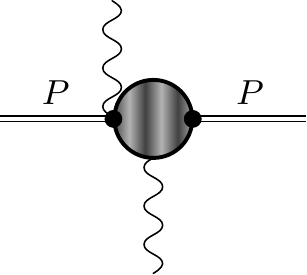}&
\includegraphics[scale=.75]{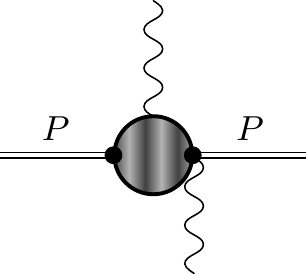}&
\includegraphics[scale=.75]{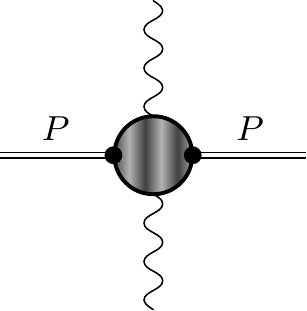}&
\includegraphics[scale=.75]{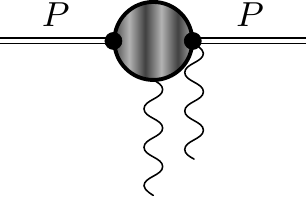}\\
(f)&(g)&(h)&(i)&(j)
\end{tabular}
\caption{The one-loop diagrams contributing to $M_{\alpha\beta}^{(a)}$. Wavy filling represents the vacuum gluon field~\eqref{b-field}. The diagrams related by crossings are not shown.\label{figure_2p2m}}
\end{figure}
The diagrams (b) and (c) contain kinematic singularities that are canceled by corresponding Born terms in Eq.~\eqref{born_scattering}.
One can notice that among these diagrams, only diagram (a) contains tensors proportional to $g_{\alpha\beta}$ and can be parametrized as
\begin{align}
\nonumber
M^{(\text{a})}_{\alpha\beta}&=\left(-\frac{t}{2}A(\nu^2,t)+\nu^2B(\nu^2,t)\right) g_{\alpha\beta}+ \text{other tensor structures},\\
\label{polarizability_from_tensor}
M^{(\text{a})\text{NB}}_{\alpha\beta}&=\left(-\frac{t}{2}A(\nu^2,t)+\nu^2B(\nu^2,t)-2\right) g_{\alpha\beta}+ \text{other tensor structures},
\end{align}
according to formulas~\eqref{born_scattering} and~\eqref{compton_tensor_parametrization}. The amplitudes $A,B$ that appear in definition of polarizabilities~\eqref{alpha_plus_beta_definition},\eqref{alpha_minus_beta_definition} can be extracted from coefficient of $g_{\alpha\beta}$ in $M_{\alpha\beta}^\text{NB}$ with the help of formulas
\begin{align*}
A^\text{NB}(0,0)&=-2\frac{\partial}{\partial t}\left.\left(-\frac{t}{2}A(\nu^2,t)+\nu^2B(\nu^2,t)-2\right)\right|_{t=0,\nu=0},\\
B^\text{NB}(0,0)&=\frac{1}{2}\frac{\partial^2}{\partial^2 \nu}\left.\left(-\frac{t}{2}A(\nu^2,t)+\nu^2B(\nu^2,t)-2\right)\right|_{t=0,\nu=0}.
\end{align*}
It is therefore sufficient to calculate only diagrams (a) of gauge-invariant combination of (a),(b) and (c) in order to extract electric and magnetic dipole polarizabilities. This is more straightforward because diagram (a) does not contain kinematic poles.
One-loop contributions of this type are shown in Fig.~\ref{figure_2p2m} (the Feynman rules in Euclidean space are given by formulas~\eqref{effective_meson_action},\eqref{Gammak} and \eqref{barG} for loops, Eqs.~\eqref{qmvert},\eqref{1-photon-vertex},\eqref{2-photon-vertex} describe non-local vertices which should be multiplied by corresponding $h_\mathcal{Q}$, the local vertices are the same as in QED, the quark propagators are defined by Eq.\eqref{quark_propagator}). For example, diagram (e) in Fig.\ref{figure_2p2m} corresponds to
\begin{equation*}
\begin{split}
(2\pi)^4\delta^{(4)}(p'+q'-p-q)M_{\mu\nu}^{\text{(a,e)}}=e^2h_P^2 &\int\! d\sigma_B\int\! d^4x \int\! d^4y \int\! d^4z \exp(ipx+iqx-ip'y-iq'z)\\
&\times(-1) \text{Tr}V_{\mu}^{aP00}(x;-q)S(x,y)V^{bP00}(y)S(y,z)Q\gamma_\nu S(z,y)\\
&+\text{crossed term},
\end{split}
\end{equation*}
for two external pseudoscalar ground-state mesons (see Appendix~\ref{section_evaluation_of_diagrams} for details).

The contribution to the amplitude corresponding to diagram (d) in Fig.~\ref{compton_scattering_diagrams} with intermediate scalar fields is separately gauge-invariant and can be represented as
\begin{equation}
\label{intermediate_scalar_meson}
\varepsilon^\alpha(q)\varepsilon^{*\beta}(q')M^{(d)}_{\alpha\beta}=\varepsilon^\alpha(q)\varepsilon^{*\beta}(q')\sum_S \Gamma_{\alpha\beta}^{S\gamma\gamma} D_S h_P^2\Gamma_{SPP},
\end{equation}
where $\Gamma_{S\gamma\gamma}$ parametrizes $S\to \gamma\gamma$ subprocess, $D_S$ is a scalar meson field propagator,
$h_P^2\Gamma_{SPP}(t)$ describes transition of scalar field to a couple of pseudoscalar mesons (corresponding one-loop diagrams are shown in Fig.~\ref{figure_intermediate_scalar}, the formulas are given in Appendix~\ref{section_evaluation_of_diagrams}). Diagram $M^{(d)}$ does not contain tensor $T_{2\alpha\beta}$ (see formula~\eqref{compton_tensor_parametrization}) and hence contributes only to amplitude $A$.
\begin{figure}
\begin{tabular}{c@{\hspace*{1em}}c@{\hspace*{1em}}c@{\hspace*{1em}}c}
\includegraphics[scale=.75]{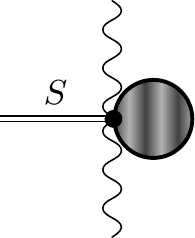}&
\includegraphics[scale=.75]{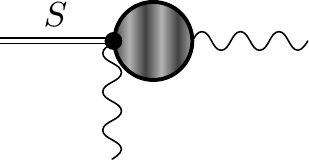}&
\includegraphics[scale=.75]{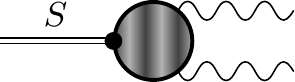}&
\includegraphics[scale=.75]{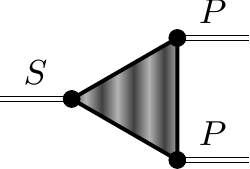}\\[1ex]
(a)&(b)&(c)&(d)
\end{tabular}
\caption{
One-loop diagrams contributing to $M_{\alpha\beta}^S$ in formula~\eqref{intermediate_scalar_meson} (crossed diagrams are not shown). Diagrams (a),(b),(c) are related to $\Gamma_{S\gamma\gamma}$, diagram (d) results in $h_P^2\Gamma_{SPP}$.
\label{figure_intermediate_scalar}}
\end{figure}

The computational complexity of the amplitudes which we need to evaluate in order to extract polarizabilities quickly grows with increasing the radial number $n$
taken into account for diagonalization of the quadratic part of the action. In the present paper, for numerical calculation   only $n=0$ states are taken into account, such that the  matrix $\mathcal{O}$ in Eq.~\eqref{physical_meson_fields} is reduced to the  unity matrix, and the higher radial states are neglected.
The values of parameters in this lowest approximation with respect to the ``radial excitation mixing'' are  given in Table~\ref{values_of_parameters} (for detailed discussion  see~\cite{Nedelko:2016gdk}).
\begin{table}
\begin{tabular}{|*{4}{@{\hspace*{0.5em}}c@{\hspace*{0.5em}}|}}
\hline
$m_{u/d}$(MeV)&$m_s$(MeV)&$\Lambda$(MeV)&$\alpha_s$\\
\hline
$174.8$&$393.3$&$439.7$&$6.23$\\
\hline
\end{tabular}
\caption{Values of parameters used for calculations of polarizabilities which were extracted from the experimental values of masses of $\pi,\rho,K,K^*$ via Eq.~\eqref{mass-eq}. 
The values of quark masses are largely affected by the presence of background gluon field, and they are not in exact one-to-one correspondence with the values of masses in models without background gluon field, see e.g. Ref.~\cite{Nedelko:2016gdk}.
\label{values_of_parameters}}
\end{table}

The values of polarizabilities found in the present work are presented in Table~\ref{polarizabilities_table}. Since no small-momentum expansion is employed, we can also calculate polarizabilities of kaons. In contrast with results obtained in several distinct quark-meson models~\cite{Ivanov:1991kw,Dorokhov:1997rv,Hiller:2009ik}, the main contribution in the model under consideration comes from one-loop diagrams, while contribution of intermediate scalar field is less important. However, this can be considered as a rearrangement of contributions because only their sum is observable.
\begin{table}
\begin{tabular}{|c@{\hspace*{1em}}c@{\hspace*{0.5em}}|@{\hspace*{0.5em}}c@{\hspace*{0.5em}}|@{\hspace*{0.5em}}c@{\hspace*{0.5em}}|@{\hspace*{0.5em}}c@{\hspace*{0.5em}}||@{\hspace*{0.5em}}c@{\hspace*{0.5em}}|@{\hspace*{0.5em}}c@{\hspace*{0.5em}}|}
\hline
&&Diagrams in Fig.~\ref{figure_2p2m}&Diagrams in Fig.~\ref{figure_intermediate_scalar}&Total&Experiment&ChPT\\
\hline
\multirow{2}{*}{$\pi^\pm$}&$\alpha_\text{E}+\beta_\text{M}$&$0.13$&0&$0.13$&$0.5\pm 0.5_\text{stat}$~\cite{COMPASS:2014eqi}&$0.16$~\cite{Gasser:2006qa}\\
&$\alpha_\text{E}-\beta_\text{M}$&$4.82$&$1.48$&$6.3$&$4.0\pm 1.2_\text{stat}\pm 1.4_\text{syst}$~\cite{COMPASS:2014eqi}&$5.7$~\cite{Gasser:2006qa}\\
\hline
\multirow{2}{*}{$\pi^0$}&$\alpha_\text{E}+\beta_\text{M}$&$0.71$&0&$0.71$&$0.98\pm 0.03$~\cite{Filkov:1998rwz}&$1.15$~\cite{Gasser:2005ud}\\
&$\alpha_\text{E}-\beta_\text{M}$&$-0.26$&$1.48$&$1.22$&$-1.6\pm 2.2$~\cite{Filkov:1998rwz}&$-1.9$~\cite{Gasser:2005ud}\\
\hline
\multirow{2}{*}{$K^\pm$}&$\alpha_\text{E}+\beta_\text{M}$&$0.41$&0&$0.41$&&\\
&$\alpha_\text{E}-\beta_\text{M}$&$1.22$&$0.47$&$1.69$&&\\
\hline
\multirow{2}{*}{$K^0,\bar{K}^0$}&$\alpha_\text{E}+\beta_\text{M}$&$0.62$&0&$0.62$&&\\
&$\alpha_\text{E}-\beta_\text{M}$&$-0.29$&$0.17$&$-0.12$&&\\
\hline
\end{tabular}
\caption{Numerical results for polarizabilities of pseudoscalar mesons in Gaussian units of $10^{-4}\ \text{fm}^3$. Column ``Diagrams in Fig.~\ref{figure_2p2m}'' corresponds to leading-order contribution of diagrams (a),(b),(c) in Fig.~\ref{compton_scattering_diagrams} given by diagrams in Fig.~\ref{figure_2p2m} which is extracted with formula~\eqref{polarizability_from_tensor}. ``Diagrams in Fig.~\ref{figure_intermediate_scalar}'' labels leading-order contribution of diagram (d) in Fig.~\ref{compton_scattering_diagrams} with intermediate scalar quark-antiquark fields given by formula~\eqref{intermediate_scalar_meson}.\label{polarizabilities_table}}
\end{table}

\section{Discussion}
We investigated dipole polarizabilities of the light pseudoscalar mesons in the framework of the nonlocal effective meson action obtained within the mean-field approach to QCD vacuum. The model described by the functional~\eqref{meson_pf_em} allows consistent treatment of various phenomena of low-energy hadronic physics: spectra of mesons, their decay constants and form-factors. Comparison of the present formalism with other approaches like Functional Renormalization Group, Dyson--Schwinger Equations, Lattice QCD and AdS/QCD is outlined in paper~\cite{Nedelko:2016gdk}.

The values of charged pion polarizabilities calculated in the present study are in agreement with COMPASS data and most recent two-loop ChPT calculation~\cite{Gasser:2006qa}. The pion mass and leptonic decay constant evaluated in the same framework earlier~\cite{Nedelko:2016gdk} agree with experimental data, and these values serve as phenomenological input for the basic Lagrangian of ChPT. The agreement with ChPT is then follows from identification of pion as pseudo Goldstone boson of broken chiral symmetry.
Moreover, an effective low-energy Lagrangian for pions can be obtained from generating functional~\eqref{meson_pf_em} if one integrates out heavier fields and performs an expansion in small momenta of pions. It is clear that such an analysis would be technically complicated, it deserves a separate investigation which would be interesting to perform, and we hope to do it in due course.

Prediction of Lattice QCD for polarizabilities depends on parameters such as the lattice volume, lattice spacings, quark masses. The value of dipole magnetic polarizability of charged pion found in paper~\cite{Luschevskaya:2015cko} with the finest lattice is $\beta_{\pi^\pm}=-2.06\pm 0.76\times 10^{-4}\ \text{fm}^3$ supports data of COMPASS collaboration~\cite{COMPASS:2014eqi}, ChPT~\cite{Gasser:2006qa} and findings of this paper. 

The distinctive feature of the present approach is that mesons are extended collective excitations of quark-antiquark and gluon fields in the confining gluon background field.
The structure of meson is encoded in the nonlocal meson-quark vertices~\eqref{qmvert} which are straightforwardly calculated. The nonlocality of meson-quark vertices leads to meson-quark-photon interactions given by Eqs.~\eqref{1-photon-vertex},\eqref{2-photon-vertex}. 
Another feature of the present approach is that intermediate scalar quark-antiquark field cannot be identified with physical light scalar meson because corresponding propagator has no pole at real momenta. Even though there is no light scalar quark-antiquark particles, the corresponding field contributes to dipole polarizabilities. As a result of these features, the contributions to polarizabilities are arranged differently from other quark-meson models~\cite{Ivanov:1991kw,Dorokhov:1997rv,Hiller:2009ik}, and the main contribution to polarizabilities in the model under consideration comes from one-loop diagrams in Fig.\ref{figure_2p2m}.

\begin{figure}
\includegraphics[scale=1]{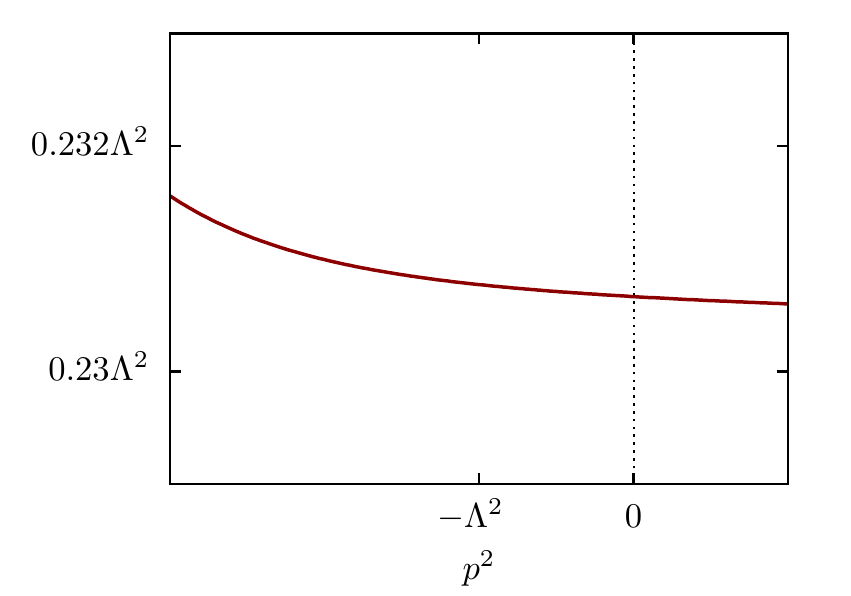}
\caption{The inverse propagator for scalar field emerging from hyperfine splitting of orbital excitation of vector meson field with respect to Euclidean momenta $p^2$. \label{figure_inverse_propagator_v1}}
\end{figure}

Besides ground-state scalar fields, the effective meson action~\eqref{meson_pf_em} contains other scalar fields. For instance, it includes the scalar component of orbitally excited vector meson field emerging from hyperfine splitting, with meson-quark vertices given by
\begin{equation*}
V^{a01n}_{\mu \mu}= \frac{1}{4}{\cal C}_{1n}\mathcal{M}^a\gamma^\mu F_{n1}\left(\frac{\stackrel{\leftrightarrow}{\cal D}^2\!\!\!
(x)}{\Lambda^2}\right)\frac{1}{i}\frac{\stackrel{\leftrightarrow}{\cal D}_\mu\!(x)}{\Lambda}.
\end{equation*}
The inverse propagator of corresponding isosinglet field in ground radial state $n=0$ at real momenta is shown in Fig.~\ref{figure_inverse_propagator_v1} 
One expects that the contribution to dipole polarizabilities of these fields via diagram (d) in Fig.~\ref{compton_scattering_diagrams} is smaller than contribution of ground-state scalar quark-antiquark field (the inverse propagator is shown in Fig.~\ref{figure_pi_propagator}) if for no other reason than their propagator is also smaller at $p^2=0$. Thorough investigation of this contribution, however, is even more complex than contribution of ground-state scalar quark-antiquark field. The zero of inverse propagator shown in Fig.~\ref{figure_inverse_propagator_v1} is located at $\sqrt{p^2}=1252-i\ 203\ \text{MeV}$. In contrast, it was found that the inverse propagator of ground-state scalar quark-antiquark field has no zeroes in complex plane in physically relevant region $\left| p^2 \right|<\left(2\ \text{GeV}\right)^2$.

The computation has been performed in the lowest approximation with respect to the mixing of radially excited states in the functional~\eqref{meson_pf_em}, and it would be interesting and important to check the stability of obtained results in this respect by accounting the higher radial excitations,
which will also allow one to estimate the polarizabilities of  radially   excited pion and kaon states. However, the latter has mostly  purely theoretical importance as experimental measurement seems to be hardly achievable. More detailed discussion about relation of the present approach to ChPT is an interesting issue which we expect to address in future work.

\section{Acknowledgements}
We are grateful for resources provided by ``Govorun'' supercomputer at the Joint Institute for Nuclear Research.

\appendix
\section{Evaluation of diagrams\label{section_evaluation_of_diagrams}}
\subsection{Formulas for diagrams in Figure~\ref{figure_2p2m}}
The notations are given in Sections~\ref{section_model} and~\ref{section_polarizabilities}. Indices $a,b$ correspond to combination of flavor matrices for a given pseudoscalar ground-state meson. The one-loop contribution to $M_{\mu\nu}^{(a)}$ in Fig.~\ref{compton_scattering_diagrams} is given by
\begin{equation*}
M_{\mu\nu}^{\text{(a)}}=\sum_{\text{k}=\text{a,b,c,d,e,f,g,h,i,j}}M_{\mu\nu}^{\text{(a,k)}}+\text{crossed diagrams}.
\end{equation*}
Here $M_{\mu\nu}^{\text{(a,k)}}$ are given by diagrams in~Fig.~\ref{figure_2p2m}:
\begin{gather*}
\tag{a}
\begin{split}
(2\pi)^4\delta^{(4)}(p'+q'-p-q)M_{\mu\nu}^{\text{(a,a)}}=
e^2h_P^2& \int\! d\sigma_B \int\! d^4x \int\! d^4y \exp(ipx+iqx-iq'x-ip'y)\\
&\times(-1) \text{Tr}V_{\mu\nu}^{aP00}(x;-q,q')S(x,y)V^{bP00}(y)S(y,x),
\end{split}\\
\tag{b}
\begin{split}
(2\pi)^4\delta^{(4)}(p'+q'-p-q)M_{\mu\nu}^{\text{(a,b)}}=
e^2h_P^2 &\int\! d\sigma_B \int\! d^4x \int\! d^4y \exp(ipx+iqy-iq'y-ip'y)\\
&\times(-1) \text{Tr}V^{aP00}(x)S(x,y)V_{\mu\nu}^{bP00}(y;-q,q')S(y,x),
\end{split}\\
\tag{c}
\begin{split}
(2\pi)^4\delta^{(4)}(p'+q'-p-q)M_{\mu\nu}^{\text{(a,c)}}=
e^2h_P^2 &\int\! d\sigma_B\int\! d^4x \int\! d^4y \int\! d^4z_1 \int\! d^4z_2\exp(ipx-ip'y+iqz_1-iq'z_2)\\
&\times(-1) \text{Tr}V^{aP00}(x)S(x,y)V^{bP00}(y)S(y,z_1)Q\gamma_\mu S(z_1,z_2)Q\gamma_\nu S(z_2,x),
\end{split}\\
\tag{d}
\begin{split}
(2\pi)^4\delta^{(4)}(p'+q'-p-q)M_{\mu\nu}^{\text{(a,d)}}=
e^2h_P^2& \int\! d\sigma_B \int\! d^4x \int\! d^4y \int\! d^4z_1 \int\! d^4z_2\exp(ipx-ip'y+iqz_1-iq'z_2)\\
&\times(-1) \text{Tr}V^{aP00}(x)S(x,z_1)Q\gamma_\mu S(z_1,z_2)Q\gamma_\nu S(z_2,y)V^{bP00}(y)S(y,x),
\end{split}\\
\tag{e}
\begin{split}
(2\pi)^4\delta^{(4)}(p'+q'-p-q)M_{\mu\nu}^{\text{(a,e)}}=
e^2h_P^2& \int\! d\sigma_B \int\! d^4x \int\! d^4y \int\! d^4z \exp(ipx+iqx-ip'y-iq'z)\\
&\times(-1) \text{Tr}V_{\mu}^{aP00}(x;-q)S(x,y)V^{bP00}(y)S(y,z)Q\gamma_\nu S(z,y),
\end{split}\\
\tag{f}
\begin{split}
(2\pi)^4\delta^{(4)}(p'+q'-p-q)M_{\mu\nu}^{\text{(a,f)}}=
e^2h_P^2& \int\! d\sigma_B \int\! d^4x \int\! d^4y \int\! d^4z \exp(ipx+iqx-ip'y-iq'y)\\
&\times(-1) \text{Tr}V_{\mu}^{aP00}(x;-q)S(x,y)V_{\nu}^{bP00}(y;q')S(y,x) ,
\end{split}\\
\tag{g}
\begin{split}
(2\pi)^4\delta^{(4)}(p'+q'-p-q)M_{\mu\nu}^{\text{(a,g)}}=
e^2h_P^2& \int\! d\sigma_B \int\! d^4x \int\! d^4y \int\! d^4z \exp(ipx+iqx-iq'z-ip'y)\\
&\times(-1) \text{Tr}V_{\mu}^{aP00}(x;-q)S(x,z)Q\gamma_\nu S(z,y)V^{bP00}(y)S(y,x),
\end{split}\\
\tag{h}
\begin{split}
(2\pi)^4\delta^{(4)}(p'+q'-p-q)M_{\mu\nu}^{\text{(a,h)}}=
e^2h_P^2& \int\! d\sigma_B \int\! d^4x \int\! d^4y \int\! d^4z \exp(ipx+iqy-ip'y-iq'z)\\
&\times(-1) \text{Tr}V^{aP00}(x)S(x,y)V_{\mu}^{bP00}(y;-q)S(y,z)Q\gamma_\nu S(z,x),
\end{split}\\
\tag{i}
\begin{split}
(2\pi)^4\delta^{(4)}(p'+q'-p-q)M_{\mu\nu}^{\text{(a,i)}}=
e^2h_P^2& \int\! d\sigma_B \int\! d^4x \int\! d^4y \int\! d^4z_1 \int\! d^4z_2\exp(ipx-ip'y+iqz_1-iq'z_2)\\
&\times(-1) \text{Tr}V^{aP00}(x)S(x,z_1)Q\gamma_\mu S(z_1,y)V^{bP00}(y)S(y,z_2)Q\gamma_\nu S(z_2,x),
\end{split}\\
\tag{j}
\begin{split}
(2\pi)^4\delta^{(4)}(p'+q'-p-q)M_{\mu\nu}^{\text{(a,j)}}=
e^2h_P^2& \int\! d\sigma_B \int\! d^4x \int\! d^4y \int\! d^4z \exp(ipx+iqz-ip'y-iq'y)\\
&\times(-1) \text{Tr}V^{aP00}(x)S(x,z)Q\gamma_\mu S(z,y)V_{\mu}^{bP00}(y;q')S(y,x).
\end{split}
\end{gather*}
The trace is taken with respect to flavor, color and spinor indices.
Crossed diagrams can be obtained by $q\leftrightarrow q',\mu\leftrightarrow \nu$. The vertex operator 
\begin{equation*}
V^{aP00}(x)=\mathcal{M}^a i\gamma_5 \int_0^1 dt\ \exp\left(\frac{\stackrel{\leftrightarrow}{\cal D}^2\!\!\!
(x)}{\Lambda^2}t\right)
\end{equation*}
is a function of $\stackrel{\leftrightarrow}{\cal D}_\mu$ which acts as
\begin{align*}
&S_f(y-x) \stackrel{\leftrightarrow}{\cal D}_\mu\!(x)\ S_{f'}(x-z)\\
=& e^{-\frac{i}{2}y_\mu  \hat{B}_{\mu\nu}x_\nu}\int \frac{d^4p}{(2\pi)^4}e^{-ip(y-x)}\tilde{H}_f(p) \stackrel{\leftrightarrow}{\cal D}_\mu\!(x)\ e^{-\frac{i}{2}x_\mu  \hat{B}_{\mu\nu}z_\nu}\int \frac{d^4q}{(2\pi)^4}e^{-iq(x-z)}\tilde{H}_{f'}(q)\\
=&\int\frac{d^4p}{(2\pi)^4}\int\frac{d^4q}{(2\pi)^4}e^{-\frac{i}{2}y_\mu  \hat{B}_{\mu\nu}x_\nu}e^{-ip(y-x)}\tilde{H}_f(p)\left\{\xi_f\left[ip_\mu-\frac{i}{2}\hat{B}_{\mu\nu}\left(x_\nu-y_\nu\right)\right]+ \xi_{f'}\left[iq_\mu-\frac{i}{2}\hat{B}_{\mu\nu}\left(x_\nu-z_\nu\right)\right]\right\}\\
&\times e^{-\frac{i}{2}x_\mu  \hat{B}_{\mu\nu}z_\nu}e^{-iq(x-z)}\tilde{H}_{f'}(q).
\end{align*}
The loop integrals are finite due to non-local meson vertices, so no regularization is needed. With meson vertices and quark propagators given by formulas~\eqref{qmvert} and~\eqref{quark_propagator}, the space and momentum integrals are Gaussian and can be computed analytically. The averaging over background field is performed with the help of formula~\eqref{averaging_over_vacuum_field} where tensor $J_{\mu\nu}$ is a combination of external momenta of mesons and photons. 

After these straightforward transformations one arrives at integrals over proper times $s_i,t_i$ that originate from vertices and propagators. These integrals are computed numerically. Unfortunately, the analytical expressions are too cumbersome to be presented here.

\subsection{Formulas for diagrams in Figure~\ref{figure_intermediate_scalar}}
The one-loop contribution to $\Gamma_{\mu\nu}^{S\gamma\gamma}$ is given by
\begin{equation*}
\Gamma_{\mu\nu}^{S\gamma\gamma}=\Gamma_{\mu\nu}^{\text{(a)}S\gamma\gamma}+\Gamma_{\mu\nu}^{\text{(b)}S\gamma\gamma}+\Gamma_{\mu\nu}^{\text{(c)}S\gamma\gamma}+\text{crossed diagrams},
\end{equation*}
where $\Gamma_{\mu\nu}^{\text{(k)}S\gamma\gamma}$ are given by diagrams (a),(b),(c) in~Fig.~\ref{figure_intermediate_scalar}. The one-loop contribution to $\Gamma^{SPP}$ is given by diagram (d) in~Fig.~\ref{figure_intermediate_scalar}:
\begin{gather*}
\tag{a}
\begin{split}
(2\pi)^4\delta^{(4)}(p'+q'-p-q)\Gamma_{\mu\nu}^{\text{(a)}S\gamma\gamma}=
e^2 &\int\! d\sigma_B \int\! d^4x \exp(ipx-ip'x+iqx-iq'x)\\
&\times(-1) \text{Tr}V_{\mu\nu}^{aS00}(x;-q,q')S(x,x),
\end{split}\\
\tag{b}
\begin{split}
(2\pi)^4\delta^{(4)}(p'+q'-p-q)\Gamma_{\mu\nu}^{\text{(b)}S\gamma\gamma}=
e^2& \int\! d\sigma_B \int\! d^4x \int\! d^4z \exp(ipx-ip'x+iqx-iq'z)\\
&\times(-1) \text{Tr}V_{\mu}^{aS00}(x;-q)S(x,z)Q\gamma_\nu S(z,x),
\end{split}\\
\tag{c}
\begin{split}
(2\pi)^4\delta^{(4)}(p'+q'-p-q)\Gamma_{\mu\nu}^{\text{(c)}S\gamma\gamma}=
e^2& \int\! d\sigma_B \int\! d^4x \int\! d^4z_1 \int\! d^4z_2\exp(ipx-ip'x+iqz_1-iq'z_2)\\
&\times(-1) \text{Tr}V^{aS00}(x)S(x,z_1)Q\gamma_\mu S(z_1,z_2)Q\gamma_\nu S(z_2,y),
\end{split}\\
\tag{d}
\begin{split}
(2\pi)^4\delta^{(4)}(p'+q'-p-q)h_P^2\Gamma^{SPP}=
h_P^2 \int\! d\sigma_B&\int\! d^4x \int\! d^4y \int\! d^4z \exp(iqx-iq'x-ip'y+ipz)\\
&\times(-1) \text{Tr}V^{aS00}(x)S(x,y)V^{bP00}(y)S(y,z) V^{cP00}(z) S(z,y).
\end{split}
\end{gather*}
The scalar two-point correlation function is an example where the final formula used for numerical computation has concise form:
\begin{align*}
\Gamma_{abS}^{(2)}(x-y)=&\int\! d\sigma_B \text{Tr}V^{aS00}(x)S(x,y)V^{bS00}(y) S(y,x),\\
\tilde\Gamma_{abS}^{(2)}(p)=&\mathrm{Tr}\left(\mathcal{M}^a\mathcal{M}^b\right)\frac{\Lambda^2}{4\pi^2}
\ \textrm{Tr}_\upsilon \int\limits_0^1\! dt_1\! \int\limits_0^1\! dt_2\! \int\limits_0^1\! ds_1\! \int\limits_0^1\! ds_2\! \left(\frac{1-s_1}{1+s_1}\right)^{m_f^2/4\upsilon \Lambda^2}\left(\frac{1-s_2}{1+s_2}\right)^{m_{f'}^2/4\upsilon \Lambda^2}\times\\
&\times \frac{1}{\Phi_2^2}\left[\frac{p^2}{\Lambda^2}\frac{F_1}{\Phi_2^2}+\frac{m_f m_{f'}}{\Lambda^2}\frac{F_2}{(1-s_1^2)(1-s_2^2)}+\frac{F_3}{\Phi_2}\right]\exp\left\lbrace -\frac{p^2}{2\upsilon \Lambda^2}\frac{\Phi_1}{\Phi_2}\right\rbrace,
\end{align*}
where
\begin{gather*}
\Phi_1=s_1s_2+2\left(\xi_{f'}^2s_1+ \xi_f^2s_2\right)(t_1+t_2)v,\\
\Phi_2=s_1 + s_2 + 2 (1 + s_1 s_2) (t_1 + t_2) \upsilon + 
 16 (\xi_{f'}^2 s_1 + \xi_f^2 s_2) t_1 t_2 \upsilon^2,\\
F_1=(1+s_1s_2)\left[2(\xi_{f'}s_1+\xi_f s_2)(t_1+t_2)\upsilon+4\xi_f \xi_{f'}(1+s_1s_2)(t_1+t_2)^2\upsilon^2+ s_1s_2(1-16\xi_f \xi_{f'}t_1 t_2\upsilon^2)\right],\\
F_2= (1 + s_1 s_2)^2,\\
F_3=4 \upsilon(1 + s_1 s_2) (-1 + 16 \xi_f \xi_{f'} t_1 t_2 \upsilon^2).
\end{gather*}
Analogous formulas for $\Gamma_{V,P}$ obtained in Ref.~\cite{Nedelko:2016gdk} describe the spectrum of radially excited mesons: light, heavy-light mesons and heavy quarkonia.
\bibliography{references}
\end{document}